=1pc

# Ground State and Excitations of Disordered Boson Systems

Michael Ma, Pornthep Nisamaneephong, and Lizeng Zhang*

*Department of Physics, University of Cincinnati*
*Cincinnati, Ohio 45221-0011*
*\*Department of Physics and Astronomy, The University of Tennessee*
*Knoxville, Tennessee 37996-1200*
*and*
*Solid State Division, Oak Ridge National Laboratory*
*Oak Ridge, Tennessee 37831*

*After an introduction to the dirty bosons problem, we present a gaussian theory for the ground state and excitations. This approach is physically equivalent to the Bogoliubov approximation. We find that ODLRO can be destroyed with sufficient disorder. The density of states and localization of the elementary excitations are discussed.*

## 1. INTRODUCTION

In 1958 Anderson[1] showed that eigenstates in a disordered medium can be localized if the disorder is sufficiently strong, a phenomenon now called Anderson localization. Subsequently, Mott pointed out the existence of the mobility edge[2], a critical energy separating localized eigenstates from extended ones. The effects of Anderson localization on a many-body fermion system have been studied. In a non-interacting fermion system, the system is either a metal or an insulator depending on whether the states at the fermi energy are extended or localized, even though the density of states remain non-zero either way. The metal-insulator transition occurs when the fermi energy and the mobility edge coincide. For interacting fermions, much understanding can be obtained by going to a fermi liquid description, and one can think of Anderson localization of quasiparticles[3]. However, the critical phenomena of the metal-insulator transition of interacting fermions, especially if spin degrees of freedom are present, remain largely an unsolved problem.

It is natural to ask the corresponding questions for a many-body boson system. Specifically, we consider one where the bosons can Bose condense at low enough temperature without disorder. Since the ground state in this case is a superfluid (or superconductor, if the bosons are charged), it opens the possibility of

## Ground State and Excitations of Disordered Boson Systems

a superfluid-insulator transition. Theoretical interest in this problem, called "dirty bosons" is further stimulated recently by experiments on two systems. The first is $^4$He absorbed in vycor or aerogels,[4] where it is seen that a critical coverage is necessary for supefluidity at zero temperature. The second is 2D superconducting films where a superconductor-insulator transition is observed with decreasing thickness[5] or increasing magnetic field.[6] While it is by no means clear that these systems can be adequately modelled by point-like bosons in a disordered potential, it is also undebatable that they contain aspects of the dirty boson problem.

Without disorder, it is well established that superfluidity of interacting bosons is intimately related to the Bose-Einstein condensation of free bosons. Of course, the interacting system differs from the non-interacting in at least one crucial way: the quadratic spectrum becomes linear. Nevertheless, free bosons is a fair starting point. With disorder, this is not the case. Without interactions, at $T = 0$ all the bosons will occupy the lowest energy state which, being a band-tail state, will be localized for any finite amount of disorder.[7] Furthermore, since the density of states of the band-tail will not exhibit the singular behavior of $d < 2$ band edges, the Bose condensation occurs at finite temperature in any dimension. However, localized states are insensitive to boundary condition[8,9], and this Bose condensed state is evidently not a superfluid. The localized condensate is unstable with respect to repulsive interactions. For an extended condensate, the repulsion energy per particle is of order 1, while for a localized condensate, it is of order $N$, which by far overwhelms the condensation energy. Thus, the non-interacting model is a poor starting point when disorder is present. Instead, a "minimal" model for dirty bosons must contain repulsive interactions between bosons in addition to the (one-body) kinetic energy and random potential:

$$\mathcal{H} = \sum_i \frac{-\nabla_i^2}{2m} + V(\mathbf{r}_i) + \frac{1}{2} \sum_{ij} u(\mathbf{r}_i - \mathbf{r}_j) \;, \tag{1}$$

In (1), both the random potential and the repulsion terms are "localizing" since interparticle repulsion restricts a particle from moving too close to other particles. But what about the interplay between the randomness and the repulsive interaction. We will see that they are in fact competing at the classical or Hartree level, but cooperative beyond that. For the interacting system without disorder, much understanding can be achieved by first applying the Hartree approximation and then the Bogoliubov theory. Let us consider the Hartree approximation in the disordered case; thus we look for ground state wavefunction of the form $\Psi(\mathbf{r}_1, ....., \mathbf{r}_N) = \prod_i \phi(\mathbf{r}_i)$. Minimizing energy results in the non-linear Schroedinger-like eq.:

$$-\frac{\nabla^2}{2m}\phi(\mathbf{r}) + V(\mathbf{r})\phi(\mathbf{r}) + N \int u(\mathbf{r} - \mathbf{r}')|\phi^2(\mathbf{r}')|d\mathbf{r}'\phi(\mathbf{r}) = \mu\phi(\mathbf{r}) \;, \tag{2}$$

The normalization here is $\int |\phi^2(\mathbf{r})|d\mathbf{r} = 1$, and so we see that because of the non-linear term, $\phi$ localized is ruled out no matter how strong the disorder is. While this does not rule out a condensate which although spans a finite fraction of space, is composed of disconnected "droplets" (solid line in Fig 1), the kinetic energy can always be lowered with basically no cost in the other terms by connecting up the droplet (dotted line in Fig 1). Thus, the Hartree approximation always yields a

M. Ma, P. Nisamaneephong, and L. Zhang

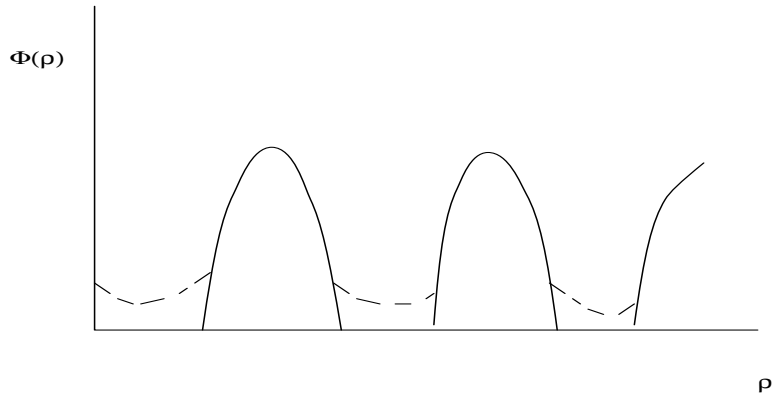

Fig. 1. The Hartree wavefunction along an arbitrary direction.

condensate which is finite everywhere and extends throughout space. This extended condensate is boundary condition sensitive due to the kinetic energy term and hence the superfluid density is finite.[8]

It is not surprising that the ground state in the Hartree approximation is always a superfluid. To destroy the superfluid, phase coherence must be destroyed. However, the random potential couples directly only to the density, and any effect on the phase must be mediated through the latter. This will happen because the local density and phase are in fact conjugate variables, obeying the commutation relationship $[\rho, \theta] = i$. But it is precisely this non-commutativity that is ignored in the Hartree approximation.

There is ample theoretical evidence that the correct solution of the dirty boson problem has a non-superfluid ground state if the disorder is strong enough[10-18]. This was shown analytically[10] by means of basically a localon expansion approach for a model of hard-core bosons on a lattice, and also in numerical calculations on finite systems on that and other models[17,18]. Indeed, we will show in this paper that a Bogoliubov theory of dirty bosons already gives a superfluidity instability.

## 2. THE SUPERFLUID AND BOSE GLASS PHASES

Thus, the generic $T = 0$ phase diagram of the dirty boson problem on a $U - W$ plot, where $U$ and $W$ are some measures of the repulsion and disorder respectively, will have two phases (we ignore the possibility of having a solid phase without disorder). The critical curve is a function of the density $n$ or the chemical potential $\mu$, and so experimentally, the transition can be probed by varying these latter parameters. As far as as the transition as a function of temperature, the critical temperature for the superfluid transition will decrease to zero at the $T = 0$ critical point.

Having established the generic dirty boson problem has a phase transition, the following are naturally issues of interest: 1) What are the two phases, ie. what

## Ground State and Excitations of Disordered Boson Systems

are the ground states and their elementary excitations.; 2) what is the critical phenomena of the $T = 0$ transition; and 3) what can the Bogoliubov theory, so important as a microscopic theory in the pure case, tell us here? We will not address the second problem here, limit ourselves to a few general statements on the first point, and after that devote the rest of this article to the last point.

The two phases are characterized by the presence or absence of off-diagonal long range order. The ground state properties of the ordered phase of dirty bosons are fundamentally equivalent to those without disorder, although the condensate is non-uniform and the superfluid density can be significantly less than the total density. The elementary excitations are not well understood. It is believed that the low-lying excitations are still phonons. In the pure case, the spectrum is linear at low momentum or equivalently at low frequency. Since momentum is not a good quantum number with disorder, one looks at the density of states at low frequency $N(\omega) \sim \omega^\theta$ and asks if it deviates from the $\theta = d - 1$ behavior without disorder. Correspondingly, the low temperature specific heat $C_v \sim T^{\theta+1}$. Another question is the Anderson localization of the phonons. The zero frequency (Goldstone) mode is always extended, since it corresponds to a uniform phase rotation (but unlike the pure case, not uniform density oscillation also). However, with sufficient disorder, the mobility edge energy can approach $0_+$, in which case there will be no propagating zero sound mode. It is interesting to ask if this localization transition is correlated with the localization transition of the ground state.

In the disordered phase, Bose condensation and superfluidity are destroyed. But is it an "insulator" or a "normal liquid", ie. is transport diffusive or activated? Since the density matrix is exponentially decaying, so is the (unaveraged) one-particle Green's function. Thus, we argue that two-particle correlation functions, including current- current correlations, must also be short-ranged; and this is an insulating phase. The low-lying excitations in this case are single-particle like, and involve the transition of a boson from one localized state to another. Because of the disordered potential, such excitations are gapless, and their density of states is finite at vanishing energy. For example, consider the extreme or "site"- localized limit, so that any overlap between localized states can be ignored. In this case, the kinetic energy term in (1) can be ignored, each localized state can be labelled by a "site" index, denoting the position of its center, and we can write the effective Hamiltonian

$$\mathcal{H} = \sum_i (\mu_i - \mu) n_i + \frac{U}{2} \sum_i n_i(n_i - 1) \ , \qquad (3)$$

where $U$ is the repulsion energy between two bosons in the same localized state, $n_i$ is the number of bosons on site $i$, and $\mu_i$ is the random local chemical potential. We assume each $\mu_i$ obeys an independent probability distribution $P(\mu_i)$, characterized by width $W$. (3) can be rewritten as

$$\mathcal{H} = \sum_i \epsilon_i n_i \ , \qquad (4)$$

where $\epsilon_i = \mu_i + U n_i$. The ground state is simply given by filling each site with the minimum $n_i$ necessary to have all $\epsilon_i > 0$. $\epsilon_i$ is the energy to add a boson to the site $i$, and is directly related to the tunneling density of states $N_1(\epsilon)$. Provided the probability of $\mu_i = 0$ is non-zero, $N_1(0)$ will be non-zero independent of the value of $U$. In particular, for $U$ large compared to $W$, $N_1(\epsilon) = P(\epsilon)$, while for $U$ small, $N_1(\epsilon)$ for $\epsilon < U$ will be enhanced by the factor $W/U$ since all sites with $\mu_i < 0$ will have $0 < \epsilon_i < U$ in the ground state. Elementary excitations correspond to the transition of a boson from an occupied site $i$ to some other (occupied or unoccupied) site $j$. Such an excitation has excitation energy $\omega_{ij} = \epsilon_j - \epsilon_i + U$. Again, the density of states $N(\omega)$ will be finite at arbitrary low energy, and with the enhancement factor for small $U$. Because $N(0)$ is finite, $C_v$ at low temperature will be linear. In the case of small $U$, a relatively sharp decrease in slope will occur at $T \sim U$. The gapless insulating phase is known in literature as the Bose glass phase.[11]

## 3. GAUSSIAN THEORY

A gaussian theory for dirty bosons will now be presented.[19] It is intrinsically equivalent to the Bogoliubov approximation. Such a theory is valuable for various reasons. For boson systems without disorder, Bogoliubov theory provides a simple microscopic foundation for more quantitative but phenomenological and/or variational theories. It corrects the quadratic excitation spectrum of the Hartree theory to the correct linear form by incorporating the zero-point fluctuations of the excitations. In a functional integral formalism, the Hartree approximation is equivalent to the saddle point approximation, and the Bogoliubov theory to including gaussian fluctuations.[20] Being a gaussian or quadratic theory, it can in principle always be solved exactly. It is hoped that the Bogoliubov theory will be a successful qualitative microscopic theory for the dirty boson problem also. For the ground state, since the Hartree approximation fails to give the insulating phase, it is of interest to investigate whether the inclusion of zero point quantum fluctuations will allow the instability of the superfluid phase with sufficient disorder. Of course, the correct critical phenomena of the superfluid- insulator transition presumably cannot be obtained by this approach. For the excitations, we hope the theory will give the correct qualitative behavior and allow us to answer the questions of density of states (and hence the specific heat) and phonon localization discussed above.

### 3.1. THE MODEL AND THE APPROACH

In principle, we can develop such a theory for the Hamiltonian (1) or its lattice version. This program has been pursued by Lee and Gunn,[21] Meng and Huang,[22] and most recently by Singh and Rokshar.[23] There are certain technical problems, including the numerical difficulty of solving the non-linear Hartree equation. Instead, we choose to study a different model, that of hard-core bosons on a lattice with on-site disorder. Our philosophy for the instability of the superfluid also differs from these references. The Hamiltonian is

$$\mathcal{H} = -\sum_i h_i b_i^\dagger b_i - J \sum_{<i,j>} (b_i^\dagger b_j + h.c.) \ , \qquad (5)$$

where the random potential $\{h_j\}$ is given by independent gaussian distribution function $P(h_j)$ with width $h$. The hard-core constraint implies $n_i$ has eigenvalues

## Ground State and Excitations of Disordered Boson Systems

0, 1. This model contains all the features we argued above to be essential for this problem, and so the results we obtain should give us general information about dirty bosons. The classical or Hartree approximation for this model corresponds to minimizing the energy with respect to Jastrow wavefunctions given by Gutzwiller projection of condensate wavefunction $|\Psi\rangle = (\sum_j \frac{v_j}{u_j} b_j^\dagger)^N |0\rangle$. The solution for $\frac{v_j}{u_j}$ is finite and has an uniform phase everywhere, no matter the value of $h$. The simplest way to obtain this result[10] is to relax the definite total number of particles $N$ constraint, and generalize to variational wavefunctions of the form $|\Psi'\rangle = \prod_j (u_j + v_j b_j^+)|0\rangle$, where the hard-core constraint is automatically satisfied. $|\Psi\rangle$ is then obtained by projecting $|\Psi'\rangle$ into a definite $N$ state. Since $<b_i> = u_i^* v_i$, ODLRO persists for all disorder in this approximation. Again, this is simply due to the disorder coupling only to the magnitude and not the phase of the order parameter.

It is convenient to use the well-known equivalence between hard-core boson operators and spin-1/2 operators to map (5) into the spin-1/2 XY magnet with a transverse random field:[10,13,14]

$$\mathcal{H} = -J \sum_{<i,j>} (S_i^x S_j^x + S_i^y S_j^y) - \sum_j h_j S_j^z \quad , \tag{6}$$

The off-diagonal LRO of the boson system is related to the magnetic LRO in the $x-y$ plane. For later convenience, we perform a global rotation of the spin-axis $x, y, z \rightarrow z, x, y$. In the spin language, the Hartree solution corresponds to treating the spins as classical vectors. Gaussian fluctuations, which take into account the zero-point motion of the spins, can be studied by applying the spin-wave theory appropriately. We propose that this approach is studying essentially the same physics as a Bogoliubov theory for Hamiltonian (1).

We now derive the spin-wave Hamiltonian.[15] First we generalize Hamiltonian (6) to arbitrary spin $S$ by rescaling $J \rightarrow J/S^2$ and $h_j \rightarrow h_j/S$. In the infinite $S$ limit, the spins behave classically. Taking the $z$-axis as the ordering axis, the spin on site $j$ lies on the $y - z$ plane at angle $\theta_j$ from the $z$-axis, with $\{\theta_j\}$ given self-consistently by

$$\sin \theta_j J \sum_{<j'>} \cos \theta_{j'} = h_j \cos \theta_j \quad , \tag{7}$$

where $<j'>$ indicates nearest neighbors of the site $j$. The statement that LRO always persists is revealed by the solution to (7) having all $\cos \theta_j \neq 0$ no matter what value of $h$ is. After a local rotation about the $x$-axis is performed, so that the spin points along the new $z$-axis, the usual Holstein-Primakoff transformation[24] of the spins into boson operators can now be defined in the rotated frame. To order $1/S$, one arrives at a quadratic Hamiltonian for the bosons:[15]

$$\mathcal{H} = -J \sum_{<i,j>} \cos \theta_i \cos \theta_j - \sum_j h_j \sin \theta_j - \frac{1}{2S} \sum_{<i,j>} \left( J_{ij} a_i^\dagger a_j + K_{ij} a_i a_j + H.c. \right) + \mathcal{O}(\frac{1}{S^{3/2}}) \quad , \tag{8}$$

where $J_{ij} = J(1 + \sin \theta_i \sin \theta_j) + \frac{h_j}{\sin \theta_j} \delta_{ij}$ and $K_{ij} = J(1 - \sin \theta_i \sin \theta_j)$, which describes gaussian fluctuations of strength $1/S$ about the classical ground state. Previously,[15] (8) has been studied perturbatively for weak disorder. In this paper, we diagonalize (8) numerically on finite-sized lattices, and will not limit ourselves to weak disorder. This will enable us to address the destruction of LRO. (8) is formally diagonalized by a Bogoliubov transformation:[25]

$$a_j = \sum_\alpha (u_{j\alpha}\gamma_\alpha + v_{j\alpha}\gamma_\alpha^\dagger) \ , \qquad (9)$$

where $\alpha$ is the eigenstate index. We have taken the $u$'s and $v$'s to be real. The $\gamma$'s are boson operators if

$$\sum_j (u_{j\alpha}u_{j\alpha'} - v_{j\alpha}v_{j\alpha'}) = \delta_{\alpha\alpha'} \ , \qquad (10)$$

and we seek the solution

$$\mathcal{H}_{SW} = E^0 + \sum_\alpha \omega_\alpha \gamma_\alpha^\dagger \gamma_\alpha \ , \qquad (11)$$

which implies the Bogoliubov equations for $u$'s and $v$'s,

$$\omega u_{j\alpha} = -\sum_{<j'>}(J_{jj'}u_{j'\alpha} + K_{jj'}v_{j'\alpha}) \ ,$$
$$\omega v_{j\alpha} = \sum_{<j'>}(K_{jj'}u_{j'\alpha} + J_{jj'}v_{j'\alpha}) \ , \qquad (12)$$

to be 'normalized' by the condition (10). For $N$ sites, this is a $2N \times 2N$ matrix equations with $2N$ eigenstates. Note that for a given solution with eigenvalue $\omega$, there is the complimentary solution $u \leftrightarrow v$, with eigenvalue $-\omega$. However, only one of these can be consistent with (10), and the other is unphysical, leaving us with $N$ physical solutions. The Goldstone mode, corresponding to uniform spin rotation about the $z$-axis in (6), is given by $u_i = v_i \propto \cos\theta_i$.

### 3.2. RESULTS

Can the zero-point fluctuations destroy the LRO of the ground state? We investigate LRO instability in 1D and 2D. Calculations in 1D are done on lattices of size 50 - 300, averaging over 500 configurations for each value of $\Delta \equiv J/h$ (please note $\Delta$ small implies strong disorder), and in 2D on $6 \times 6$ to $11 \times 11$ lattices averaging over 200 configurations. Possible signature of instability are 1) a diverging fluctuation in the order parameter (diverging condensate depletion) as $N \to \infty$, 2) negative excitation energies, or 3) complex excitation energies (e.g., Bogoliubov solution to bosons with attractive interactions). Indeed, even in the pure case, signature 1) is observed in 1D. In the present calculation, instability criteria 2) and 3) are not observed, leaving 1) as the sole possibility. Within the spin wave approximation as formulated, the relevant quantity is

$$\delta m = \frac{1}{N}\sum_j \cos\theta_j \delta\langle S_j^z\rangle = \frac{1}{N}\sum_j \sum_{\alpha\neq 0}\cos\theta_j v_{j\alpha}^2 = \int d\omega N(\omega)v^2(\omega) \ , \qquad (13)$$

where $N(\omega) = \frac{1}{N}\sum_\alpha \delta(\omega - \omega_\alpha)$ is the density of states (DOS).

## Ground State and Excitations of Disordered Boson Systems

As remarked earlier, in 1D $\delta m$ diverges as $N \to \infty$ even without disorder. However, more precisely, $\delta m \propto \ln N$, and we view this as an indication of the known algebraic LRO of the 1D XY model,[26] hence the ground state is still a superfluid.[27] Thus, we argue that a transition with disorder is still possible, and is marked by $\delta m$ diverging faster than $\ln N$. This is in fact seen in our calculation, and is shown in Fig.2, with the critical value of $\Delta = \Delta_c \approx 0.6$ in the present model.

In 2D, $\delta m/m$ is finite as $N \to \infty$ in the pure case, which, consistent with the exact result for $S = 1/2$,[28] we take to mean the LRO is stable. Fig. 3 shows $\delta m/m$ vs. $\ln N$ for different values, and we see that for weak disorder, the LRO remains stable. With strong enough disorder, $\delta m/m$ diverges as $N \to \infty$. This transition occurs between $\Delta = 0.1$ and $\Delta = 0.08$.

It is of interest to ask whether the transition is due to a change in the DOS ($N(\omega)$) or the nature of the excitations ($v^2(\omega)$) or both. In the pure case, $v^2(\omega) \propto \frac{1}{\omega}$ for small $\omega$, while $N(\omega) \propto \omega^{d-1}$ for small $\omega$. For the infinitely strong disorder ($J = 0$) case (the large $U$ site-localized limit discussed earlier), the excitations are single spin flips, with excitation energies $|h_j|$. Hence $N(\omega)$ is simply given by the distribution of $h_j$, and is finite at low energies. It seems reasonable to expect therefore $N_0 = N(\omega \to 0)$ is finite in 1D for all $\Delta$, and the transition must be due to $v^2(\omega)$ diverging faster than $1/\omega$. This picture is confirmed by our numerical calculations and in Fig. 4 we show the DOS beyond the transition in 1D. While there is some ambiguity in deciding $N_0$ for infinite system from a finite-sized calculation, we have checked to see that the scaling of $N_0$ with $N$ is in fact consistent with a non-zero DOS at zero energy. In 2D, the ordered phase should be characterized by $N(\omega) \propto \omega$ and the disordered phase by $N_0$ finite. Our results are consistent with this. For $\omega \geq 0.1$, $N(\omega)$ is linear in $\omega$, with the slope increasing with decreasing $\Delta$. For $\Delta \leq 0.08$, $N_0$ is finite. Unfortunately, we cannot say for certain whether the DOS transition exactly occurs at the order parameter transition due to the inability of pinpointing $\Delta_c$

These excitations can be extended or localized. It is of interest to ask if their localization transition is related to the 'localization' of the ground state. It is also of interest by itself as an Anderson localization problem of the eigenstates of (8). Since the zero-mode corresponds to uniform phase rotation, it must be extended. One thus expects that possibly for a given $\Delta$, a transition from extended to localized states with increasing energy at a mobility edge energy $E_c$. Is $E_c \to 0$ correlated to $\Delta \to \Delta_{c+}$? We believe the answer is no. First, perturbatively in the disorder, the phonon mean free path is found to diverge as $E^{-(d+1)}$, and so is finite for all finite $E$.[15] Common wisdom has it that in 1D, the localization length and the mean free path are essentially identical, since any scattering is backscattering. Second, while the way the disorder enters in (8) is rather complicated, the low-lying excitations are phonons, ie.density fluctuations; and the disorder couples directly to the density. Thus, the problem should be equivalent to other well-studied phonon localization models, where it is well established that all states are localized in 1D for any disorder.[29] Hence, we believe $E_c = 0$ for any disorder in 1D. In 2D, on less firm grounds, previous work on phonon localization indicates $E_c = 0$ for arbitrarily weak disorder also.[30] Presumably, in 3D $E_c \to 0$ at some finite disorder which is unrelated to $\Delta_c$. The results of a direct calculation using the participation ratio as the indication of localization is reported elsewhere.[19]

M. Ma, P. Nisamaneephong, and L. Zhang

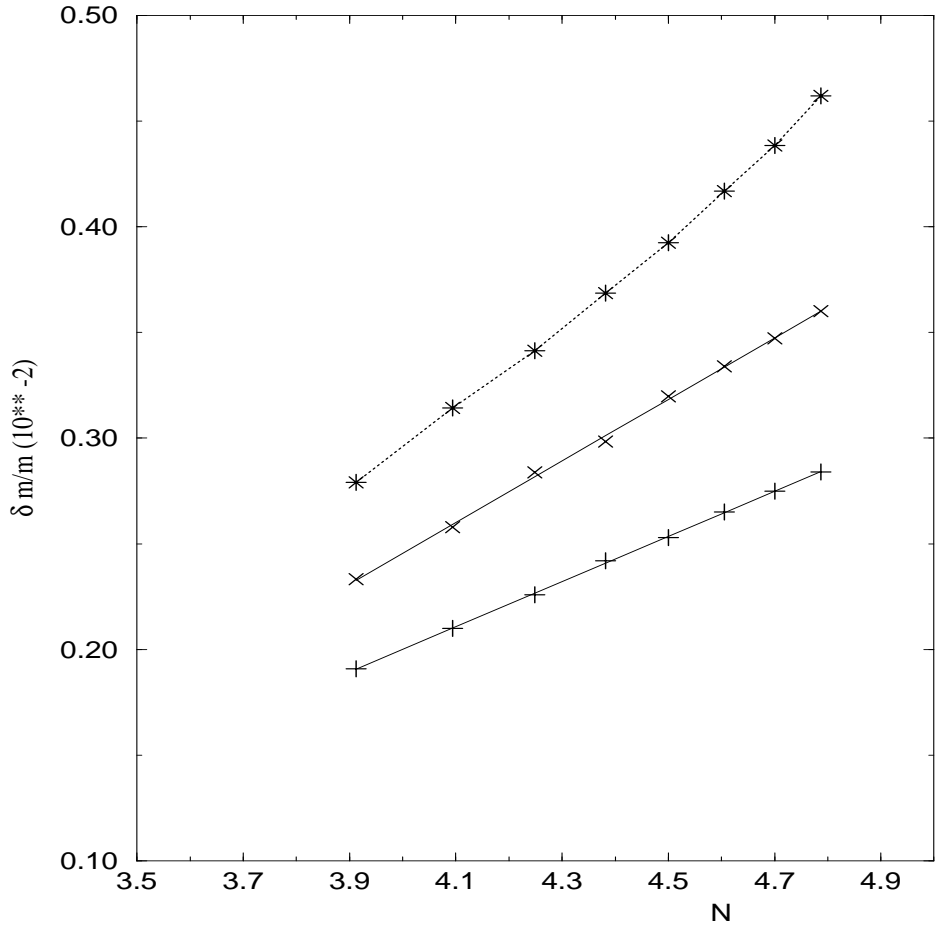

Fig. 2. Fluctuation corrections to the order parameter $\delta m/m$ is plotted against $\ln N$ in 1D. The divergence becomes faster than $\ln N$ for $\Delta$ smaller than a critical value $\Delta_c \approx 0.6$. The solid lines are obtained through a linear fit and the dashed line is a guide to the eyes.

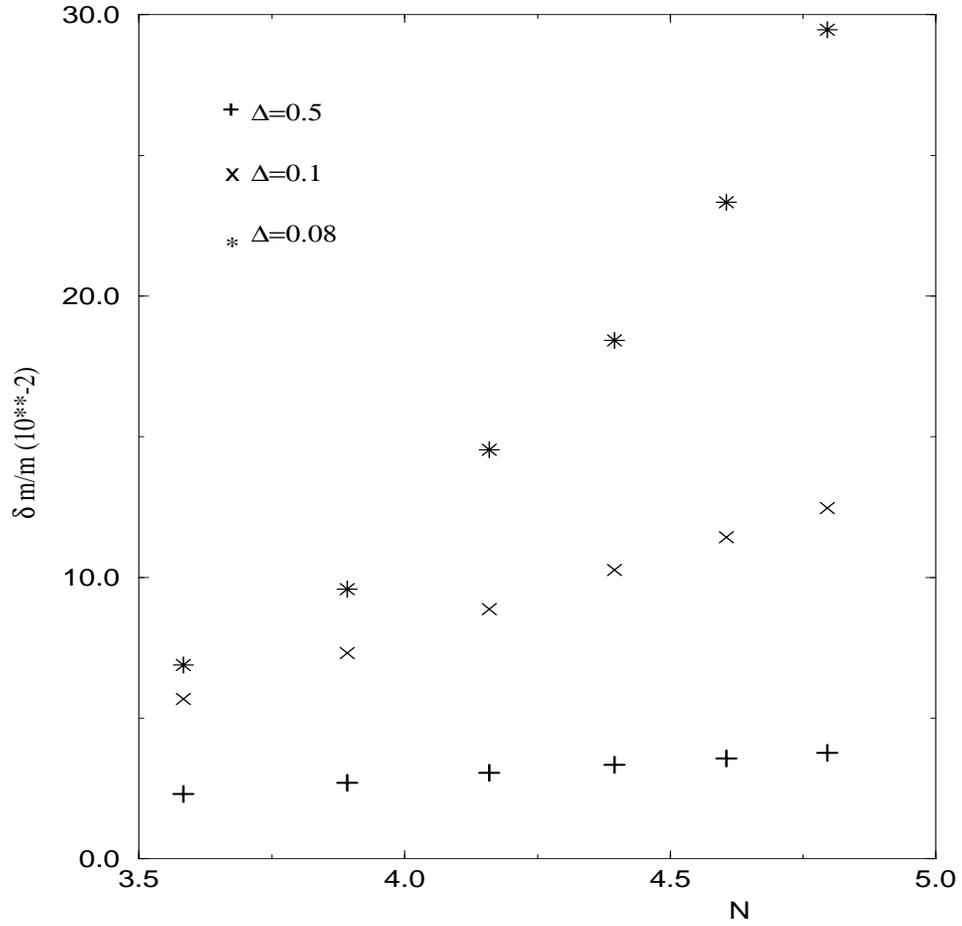

Fig. 3. The same as Fig. 2, plotted for 2D systems. Unlike it in 1D, $\delta m$ is finite for weak disorder and diverges for $\Delta < \Delta_c$, which is between 0.1 and 0.08.

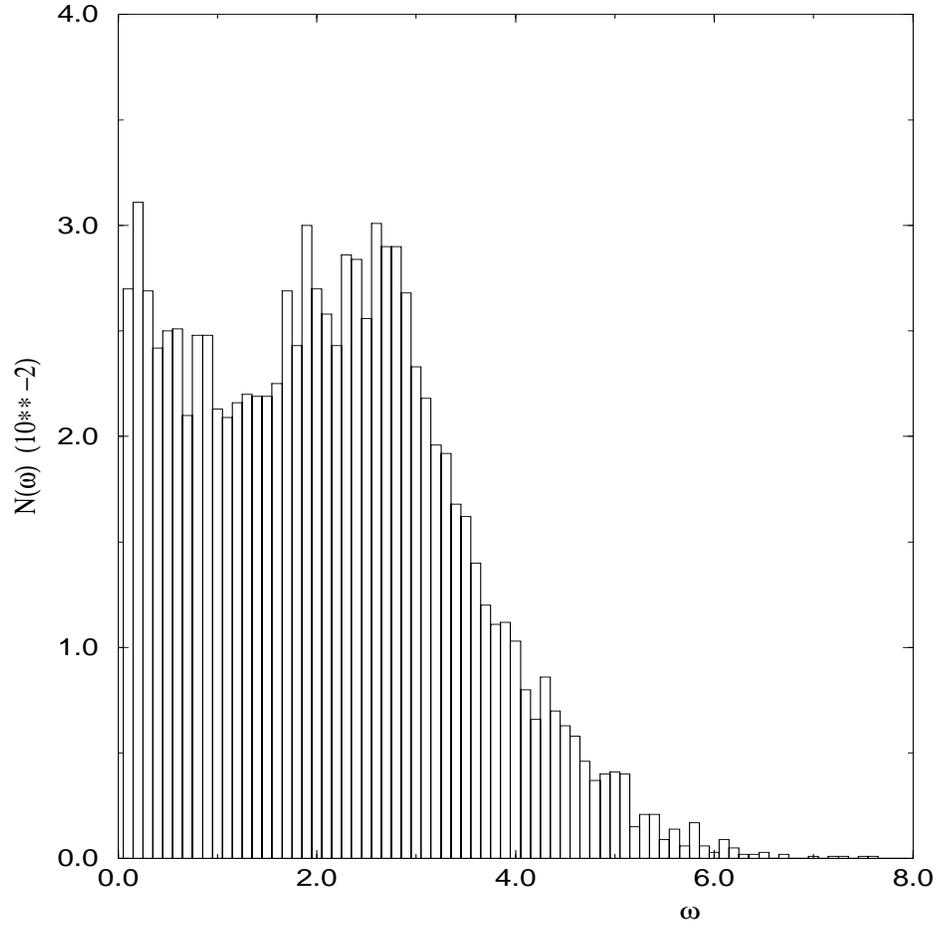

Fig. 4. DOS on the insulating side ($\Delta = 0.5$) in 1D. Here $N = 100$..

### 3.3. CONCLUSION

We have shown that a Bogoliubov-like theory for dirty bosons indicates the superfluid phase can become unstable, presumably with respect to the Bose glass phase. In our model, the critical disorder is finite and independent of the value of $1/S$, hence of the strength of gaussian fluctuations, so long as it is non-zero. We suggest the critical disorder is finite for infinitesimal repulsion in Hamiltonian (1) also. Deep in the superfluid phase, we find the low-frequency DOS of excitations to have the same energy ($N(\omega) \propto \omega^{d+1}$) dependence as pure bosons, and hence $C_v \sim T^d$. Although the theory breaks down beyond the instability point, the trend as the transition is approached suggests the Bose glass phase is gapless, which is consistent with the more general arguments presented earlier in this paper. Thus, $C_v \sim T$. In terms of using the 3D dirty boson model as the model for He in vycor, these results are not in agreements with the latest experiments which observe $C_v \sim T^2$ and $C_v$ energy gap respectively.[31] Further investigations on the applicability of the dirty boson model and its behavior are certainly warranted.

Connections can be made between Bogoliubov theory and correlated wavefunctions[32] We have seen that a Gutzwiller correlated factor cannot destroy the ODLRO. Presumably this result holds true even for longer but finite- ranged Jastrow factors. At the same time, we know long- ranged correlations are necessary to produce the linear phonon spectrum in the pure case. It would be interesting to see if a complete theory of Jastrow wavefunctions can describe the superfluid-Bose glass transition.

### ACKNOWLEDGEMENTS

We acknowledge the use of the Ohio Supercomputer. LZ acknowledges support by the National Science Foundation under Grant No. DMR-9101542 and by the U.S. Department of Energy through Contract No. DE-AC05-84OR21400 administered by Martin Marietta Energy Systems Inc.. MM thank the hospitality of the Hong Kong University of Science and Technology and the Aspen Center for Physics where portions of this work were done.